\documentclass[twocolumn,twoside,slac_two]{revtex4}
\usepackage{graphicx}
\usepackage{fancyhdr}
\newcommand{\etal}{{\it et~al.}}
\newcommand\GeV{\ifmmode {\mathrm{\ Ge\kern -0.1em V}}\else\textrm{Ge\kern -0.1em V}\fi}%

\begin{document}

\title{Measurements of the Running of the Electromagnetic Coupling at LEP}

\author{Salvatore Mele}
\affiliation{INFN, Sezione di Napoli, I-80125, Napoli, Italy}

\begin{abstract} 
The study of Bhabha scattering at $\rm e^+e^-$ colliders probes the
running of the electromagnetic coupling. After early 
measurements by the VENUS collaboration at TRISTAN and the by L3
collaboration at LEP, two recent analyses have been performed by the
OPAL and L3 collaborations. The OPAL collaboration studied
high-statistics low-angle Bhabha scattering at LEP, achieving a precise determination
of the running of $\alpha$ in the region
$1.8\GeV^2<-Q^2<6.1\GeV^2$. The L3 collaboration investigated
high-energy large-angle Bhabha scattering to
first probe the region $1800 \GeV^{2} < -Q^{2} <
21600 \GeV^{2}$. All measurements are described and a global
overview of their agreement with QED predictions is given.
\end{abstract}

\maketitle

\thispagestyle{fancy}

\section{THE RUNNING OF  {\boldmath $\alpha$}}

\begin{figure}
    \includegraphics[width=0.4\textwidth]{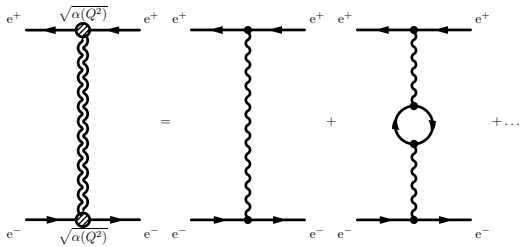}
    \caption{$t$-channel Feynman diagrams contributing to Bhabha
           scattering and the phenomenon of vacuum-polarisation.  The
           sum of all diagrams including zero, one, two or more
           vacuum-polarisation insertions is denoted by the diagram to
           the left with the double-wavy photon propagator, with an
           electromagnetic coupling $\alpha(Q^2)$.}
  \label{fig:1}
\end{figure}

A fundamental idea in quantum field theory is that coupling constants
which describe the strength of a given process are not actually
constant but rather depend on the energy scale at which the process
occurs; in other words, the constants {\it run}. In QED this running implies the
increase of the electromagnetic coupling, $\alpha$, with the squared
four-momentum transfer, $Q^2$. Figure~\ref{fig:1} sketches the origin
of this phenomenon in the case of Bhabha scattering,  $\rm e^+e^-\rightarrow
e^+e^-$: larger momentum transfers probe virtual-loop
corrections to the photon propagator.

The running of $\alpha$ is parametrised as~\cite{ref:running}:
\begin{equation}\label{eq:alfa}
  \alpha(Q^{2}) = \frac{\alpha_0}{1-\Delta\alpha(Q^{2})},
\end{equation}
where $\alpha_0$ is the fine-structure constant. This is measured with
high accuracy in solid-state processes and via the study of the
anomalous magnetic moment of the electron to be~\cite{ref:codata}:
\begin{displaymath}
 1/\alpha_0 =
137.03599911\pm0.00000046
\end{displaymath}
The term $\Delta\alpha(Q^{2})$ describes the
running of $\alpha$ in Equation~\ref{eq:alfa}. It 
receives contributions
from three sources: lepton loops, top-quark loops and loops involving
the five lighter quark flavours. At the scale of Z-boson mass, $m_{\rm
  Z}$, the
first two contributions to $\Delta\alpha(Q^{2})$ are precisely calculated as:
\begin{equation}
\Delta\alpha_{\rm e\mu\tau}(m_{\rm Z}^{2}) = 0.03150
\end{equation} 
\begin{equation}
\Delta\alpha_{\rm top}(m_{\rm Z}^{2}) = -0.00007 \pm 0.00001,
\end{equation} 
as discussed in References~\cite{ref:leptons} and~\cite{ref:top}, respectively.
The contribution from the five lighter quarks, $\Delta\alpha^{(5)}_{had}$,
is difficult to calculate due to
non-perturbative QCD effects and is estimated using
dispersion-integral techniques as~\cite{ref:burkhardt_new}:
\begin{equation}
\Delta\alpha^{(5)}_{had}(m_{\rm Z}^{2}) = 0.02758 \pm 0.00035.
\end{equation} 
The electromagnetic coupling at the scale of the Z-boson
mass is therefore~\cite{ref:burkhardt_new}:
\begin{equation}\label{eq:alphabolek}
1/\alpha(m_{\rm Z}^{2}) = 128.940 \pm 0.048.
\end{equation}

The study of fermion-pair production at LEP, $\rm e^+e^-\rightarrow
Z\rightarrow f\bar{f}$, allows a precise measurement of the Z-boson
couplings. These couplings are sensitive to electroweak
vacuum-polarisation effects and therefore also to the running of
$\alpha$. The quantity most sensitive to the running of $\alpha$ is
the effective vector coupling constant of Z bosons and electrons. It
is measured as $g_{\rm Ve}=-0.03816\pm 0.00047$~\cite{ref:smreport}.
In the absence of the running of $\alpha$ its value would be $g_{\rm
Ve}=-0.076$. A global fit to several Standard Model observables allows
to constrain the value of $\alpha$ at the scale of the Z-boson mass
as~\cite{ref:smreport}:
\begin{equation}
1/\alpha^{fit}(m_{\rm Z}^{2}) = 128.937 \pm 0.047,
\end{equation}
which is in remarkable agreement with the expected value of
Equation~\ref{eq:alphabolek}. However, this determination of $\alpha(m_{\rm Z}^{2})$
is based on the assumption of the full electroweak theory and  a determination
which relies on less hypotheses is desirable. The
study of Bhabha scattering, discussed in the following, allows a more
direct insight in the running of $\alpha$.

\section{BHABHA SCATTERING AND THE RUNNING OF {\boldmath$\alpha$}}

The study of Bhabha scattering at $\rm e^+e^-$ colliders offers a unique window
on the vacuum-polarisation 
insertions depicted in Figure~\ref{fig:1}. Its measurement gives access to the
running of $\alpha$ in the {\it space-like} region, $Q^2<0$. The
experimental conditions at $\rm e^+e^-$ colliders are extremely favourable as they allow a
precise determination of the 
four-momentum transfer through its dependence on the squared centre-of-mass energy,
$s$, and on the scattering angle, $\theta$:
\begin{displaymath} 
Q^{2} = t \simeq -s(1-\cos\theta)/2. 
\end{displaymath}
Indeed, at  $\rm e^+e^-$ colliders, $s$ is
accurately known and detectors are conceived for a precise measurement
of $\theta$. In the following, Bhabha scattering is studied in
two angular regions:
low-angle, $\theta\sim 1.5^\circ-3.5^\circ$, and large-angle, $\theta\sim
20^\circ-90^\circ$. These angular regions correspond to two energy regimes:
low-$Q^2$ and high-$Q^2$, respectively. 

The relation between the differential cross section of Bhabha scattering and the
electromagnetic coupling is:
\begin{equation}\label{eq:sigma}
  {{\rm d}\sigma \over {\rm d}t} = {{\rm d}\sigma^0 \over {\rm
  d}t} \left( {\alpha(t) \over \alpha_0}\right )^2
  (1+\varepsilon)(1+\delta_\gamma)+\delta_{\rm Z},
\end{equation}
where the tree-level cross section is:
\begin{equation}
 {{\rm
  d}\sigma^0 \over {\rm d}t} = { 4 \pi \alpha_0^2 \over t^2 }.
\end{equation}
The  $s$-channel contributions, $\delta_\gamma$ and $\delta_{\rm Z}$,
are much smaller than
the radiative corrections, $\varepsilon$. These three quantities
are known with a good accuracy compared to the experimental precision
of the study of low- and large-angle Bhabha scattering.
Na\"ively, one could imagine to measure the cross section for Bhabha
scattering and then insert the result in the
left-hand side of Equation~\ref{eq:sigma} in order to  extract a value for
$\alpha(t)$. This argument is unfortunately flawed: the measurement
of the cross section requires knowledge of the integrated luminosity.
At LEP this is estimated by counting events from low-angle Bhabha
scattering and assuming the cross section of this process to be known,
what can only happen if $\alpha(t)$ is known to start with: a {\it catch-22}~\cite{catch22}!

As discussed above, the information on $\alpha(t)$ contained in the
scale of the differential cross section  ${{\rm d}\sigma / {\rm
    d}t}$ cannot be directly unlocked. On the other hand, the shape of this
differential cross section contains information on the running of
$\alpha(t)$ over the $Q^{2}$ range in which the cross section is
measured. QED predicts that  $\alpha(t)$ increases with $t$: if the running is slower
than the one expected in QED, the differential cross section will be
steeper than expected; if the running is faster than expected in QED,
the differential cross section will be flatter than expected. 
The strategy to access information on the running of $\alpha$ is
therefore to analyse the shape of the differential cross section over
a $Q^{2}$ range and compare it with the expectation from QED. Five
such analyses exist and are presented below: we first discuss three
early measurements of the running of 
$\alpha$ at TRISTAN and LEP,
and then describe two recent measurements  performed by the
OPAL and L3 collaborations by investigating low-angle and large-angle
Bhabha scattering at LEP.

\section{EARLY MEASUREMENTS AT TRISTAN AND LEP}

LEP experiments were equipped with low-angle calorimeters, called
luminosity monitors,  engineered
to detect events from low-angle Bhabha scattering in order to measure
the integrated luminosity.  The event counts in
different polar-angle regions of the luminosity monitors depend on the
shape of the differential cross section of Bhabha scattering and,
through Equation~\ref{eq:sigma}, give access to the running of
$\alpha(t)$.

The first study of the running of $\alpha$ using low-angle Bhabha
scattering was performed by the L3 collaboration. A total of about 7 million events
collected at $\sqrt{s}=m_{\rm Z}$ in the luminosity monitor,
covering a polar angle $1.8^\circ < \theta < 3.1^\circ$, were used to 
probe the running of $\alpha$  in the region
$2.1\GeV^2<-Q^2<6.2\GeV^2$~\cite{l3-197}. 

\begin{figure}[h]
    \includegraphics[width=0.4\textwidth]{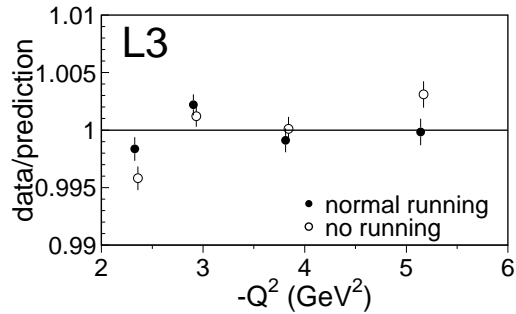}
    \caption{Ratio of measured and expected events in four
      different angular regions of the L3 luminosity monitor,
      corresponding to four values of $-Q^2,$ for a
      subset of data collected at the Z-boson pole. The solid circles
      correspond to the QED prediction and the open circles to the
      hypothesis of no running of the electromagnetic coupling. The
      data favours the QED predictions and excludes the no-running
      hypothesis. From Reference~\cite{l3-197}.}
  \label{fig:l3old}
\end{figure}

Figure~\ref{fig:l3old}
presents the ratio between the observed number of events and the QED
prediction, as well as the hypothesis of a constant value of $\alpha$
over the entire $Q^2$ range. This ratio of data and predictions is
fitted to determine the running of $\alpha$ between the two extremes
of the $Q^2$ range as:
\begin{equation}\label{eq:l3old1}
  \alpha^{-1}(-2.1 \GeV^{2}) - \alpha^{-1}(-6.2 \GeV^{2}) = 0.78 \pm
  0.26,
\end{equation}
where the uncertainty combines statistical and systematic
uncertainties. This result is in good agreement with the QED
prediction of 0.56 and first established the running of $\alpha$
in this energy range, excluding the no-running hypothesis at 4.5
standard deviations.

The running of $\alpha$ in large-angle Bhabha scattering was first
investigated by the VENUS Collaboration at TRISTAN. They studied
events collected at $\sqrt{s}=57.8\GeV$ in the angular range $47^\circ
< \theta < 90^\circ$ to cover the $Q^2$ range
$100\GeV^2<-Q^2<3337\GeV^2$~\cite{ref:venus}. 

\begin{figure}
    \includegraphics[width=0.4\textwidth]{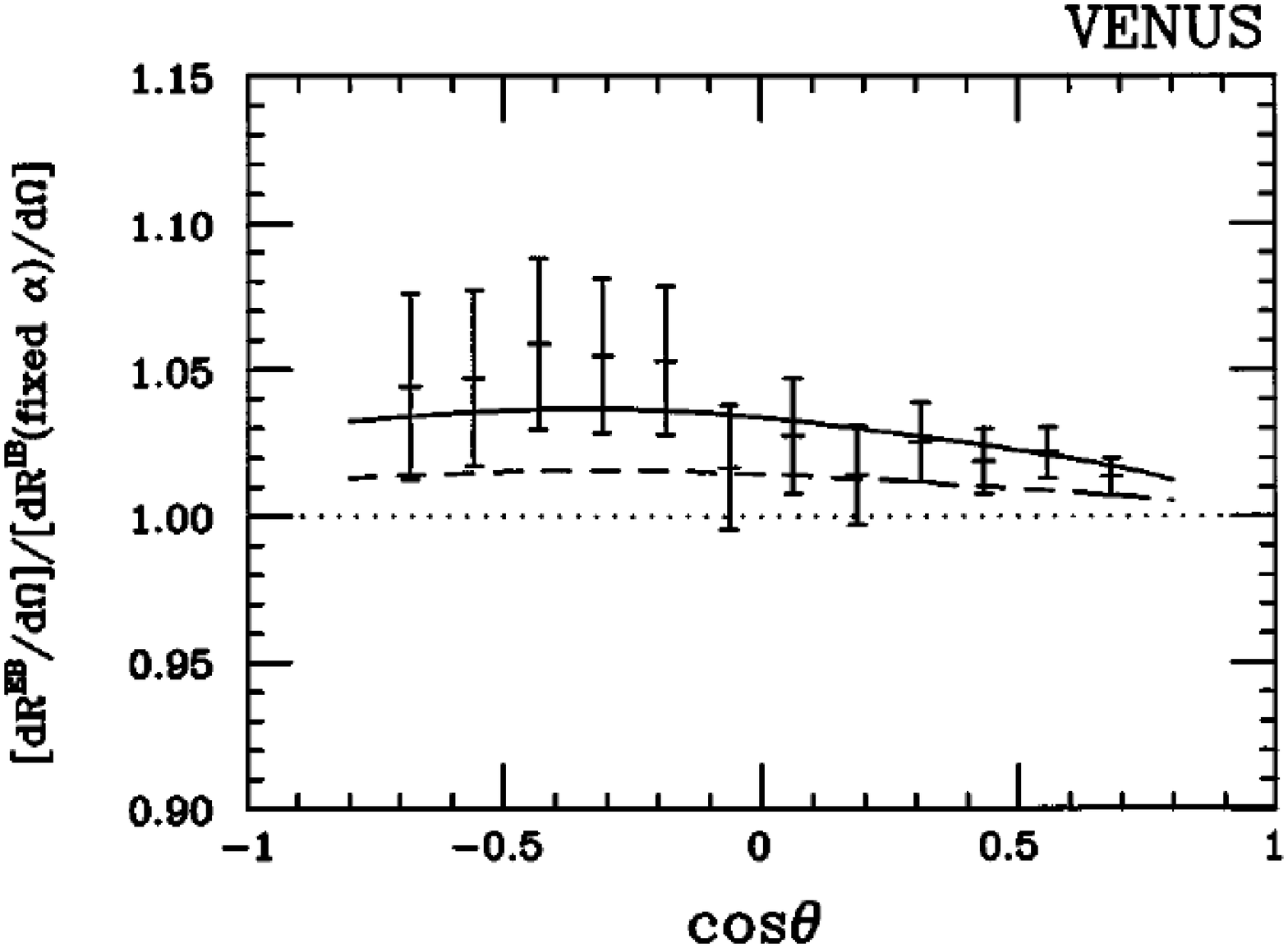}
    \caption{Ratio of the differential cross section for
      Bhabha scattering measured by the VENUS collaboration
      and the predictions for a fixed value of $\alpha$. The solid
      line shows the QED prediction for a running $\alpha$ and the
      dashed line the prediction if the running is only due to lepton
      corrections. The data favours the QED predictions. From Reference~\cite{ref:venus}.}
  \label{fig:venus}
\end{figure}

Figure~\ref{fig:venus}
shows the measured differential cross section as a function of the
cosine of the scattering angle, normalised to the QED predictions with
a fixed value of $\alpha$. The effects of the running are evident and
the data favour a scenario in which both the hadronic and leptonic
contributions are present in the running, as opposed to a scenario with just
lepton loops. The results can be expressed as the change over the
$Q^2$ range of the term
 $\Delta\alpha$ describing the running of $\alpha$ in Equation~\ref{eq:alfa}:
\begin{eqnarray}
  \Delta\alpha(-3337 \GeV^{2}) - \Delta\alpha(-100 \GeV^{2}) =\nonumber\\ 0.0186 \pm
  0.0059,
\end{eqnarray}
where the uncertainty includes statistical and systematic
effects.

A similar study at higher centre-of-mass energies and lower polar
angles was performed by the L3 Collaboration.  Bhabha-scattering
events are selected at $\sqrt{s}=189\GeV$ for scattering angles
$20^\circ <\theta<36^\circ$, probing the intermediate-$Q^2$ range
$12.25\GeV^2<-Q^2<3434\GeV^2$~\cite{l3-197}. The measured differential
cross section as a function of the scattering angle is compared to the
predictions as a function of $\alpha$, allowing to measure its running
over the $Q^2$ range as:
\begin{equation}\label{eq:l3old2}
\alpha^{-1}(-12.25 \GeV^{2}) - \alpha^{-1}(-3434 \GeV^{2}) = 3.8 \pm
1.3,
\end{equation}
where the uncertainty comprises statistical and systematic
sources. This value is in good agreement with the QED prediction of 4.07.

\section{PRECISION MEASUREMENT AT LOW {\boldmath$Q^2$}}

The OPAL collaboration analysed low-angle Bhabha scattering events to
study the running of $\alpha$ at low $Q^2$~\cite{ref:opal}. Data
collected at  $\sqrt{s}=m_{\rm Z}$ with the
luminosity monitor, consisting of layers of tungsten absorber and
32-pad silicon detectors, were used. This detector covered a polar
region $1.4^\circ <\theta<3.3^\circ$, corresponding to a momentum
transfer $1.8\GeV^2<-Q^2<6.1\GeV^2$.

\begin{figure}[h]
      \includegraphics[width=0.4\textwidth]{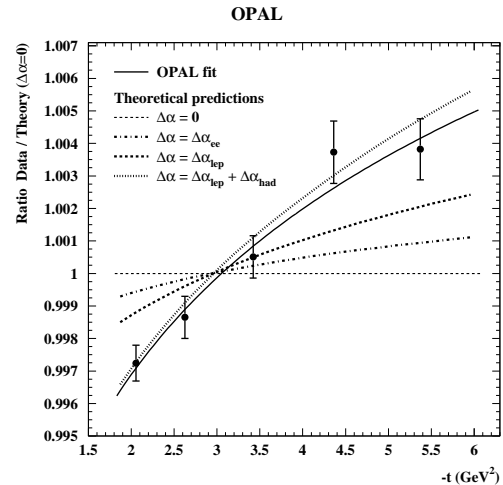}
    \caption{The ratio of event counts five
    radial regions of the OPAL luminosity monitor,
    corresponding to five $Q^2$ values, for data and a Monte Carlo simulation
    with $\alpha(t)=\alpha_0$.  The
    data favour the QED prediction for the running of $\Delta\alpha(t)$ and exclude
    scenarios with no running or a running due only to electron or
    lepton virtual loops. Only statistical uncertainties are shown. From Reference~\cite{ref:opal}.}
    \label{fig:opal}
\end{figure}

The analysis selects 10 million high-energy back-to-back cluster pairs
and studies their $t$ spectrum to extract information on
$\alpha(t)$. Data are divided in five $t$ bins and are compared with
the BHLUMI Monte Carlo~\cite{ref:bhlumi}.  The results of the study
are quantified by studying the ratio of the $t$ spectrum observed in
data and the Monte Carlo spectrum for the hypothesis
$\alpha(t)=\alpha_0$, as shown in Figure~\ref{fig:opal}.

The data in Figure~\ref{fig:opal} are fitted with the function $a+b\ln(t/t_0)$, where
$t_0=-3.3\GeV^2$ is the mean value of $t$ in the data sample.  The
parameter $b$ is related to the running of $\alpha$ by:
\begin{equation}\label{eq:b}
\Delta\alpha(t_2)-\Delta\alpha(t_1)\approx {b\over 2} \ln {t_2 \over
t_1}.
\end{equation}
Four hypotheses are tested by the OPAL analysis: no running of $\alpha$; a running only due
to electron virtual-loops; a running only due to lepton virtual-loops;
the QED hypothesis of a running induced by virtual loops of leptons
and quarks. The first three hypotheses are found to be completely
excluded by data.

The observed running of $\alpha$ is derived from the fitted value of
$b$ and Equation~\ref{eq:b} to be:
\begin{eqnarray}\label{eq:opal}
\Delta\alpha(-6.1\GeV^2)-\Delta\alpha(-1.8\GeV^2) =\nonumber\\ (440 \pm 58 \pm 43
\pm 30) \times 10^{-5},
\end{eqnarray}
where the first uncertainty is statistical, the second systematic and
the third theoretical. The main sources of systematic uncertainties
are the simulation of the detector material and the reconstruction
of the radial coordinates of the energy clusters in the
calorimeter. This results is in excellent agreement with the QED
prediction of $460 \times 10 ^{-5}$.

This measurement proves the running of $\alpha$ at low $Q^2$ with a
significance of $5.6\sigma$. In addition, it establishes for the first
time the hadronic contribution to the running with a significance of
$3.0\sigma$ as:
\begin{eqnarray}
\Delta\alpha_{had}(-6.07\GeV^2)-\Delta\alpha_{had}(-1.81\GeV^2) =\nonumber\\ (237
\pm 58 \pm 43 \pm 30) \times 10^{-5}.
\end{eqnarray}

\section{FIRST MEASUREMENT AT LARGE {\boldmath$Q^2$}}

\begin{figure}
      \includegraphics[width=0.4\textwidth]{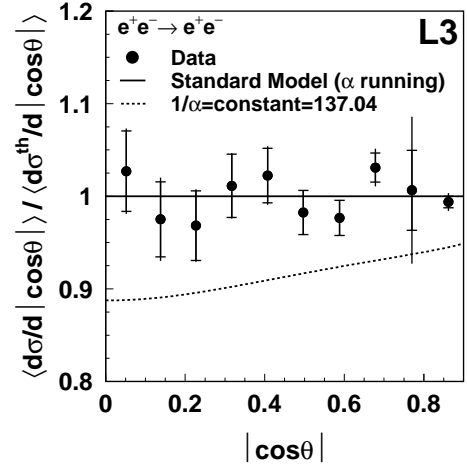}
    \caption{
    The ratio of the centre-of-mass-averaged differential cross section of
    for large-angle Bhabha scattering measured by L3 and the corresponding
    theoretical predictions. The inner error
    bars denote the statistical uncertainties, the outer the
    combination of statistical and systematic uncertainties. The data
    exclude the scenario $\alpha(t)=\alpha_0$, represented by the
    dotted line. From Reference~\cite{ref:l3}.}
    \label{fig:l3}
\end{figure}

The L3 collaboration probed the running of $\alpha$ at large $Q^{2}$
by studying large-angle Bhabha scattering in the angular region
$26^\circ < \theta < 90^\circ$~\cite{ref:l3}. Data collected at eight centre-of-mass energies
in the range $\sqrt{s}=189-209\GeV$ were
considered, corresponding to a four-momentum transfer $ 1800
\GeV^{2} < -Q^{2} < 21600 \GeV^{2}$.
About 40\,000 events are selected as back-to-back
clusters in the high-resolution BGO electromagnetic calorimeter with
matched tracks. These events are used to measure the differential cross
section of Bhabha scattering in ten different angular ranges, for each
of the eight energy points, for a
total of 80 measurements. The measured cross section are compared to
the  predictions of the BHWIDE Monte
Carlo~\cite{ref:BHWIDE} to extract information on the running of
$\alpha$.  As an example, Figure~\ref{fig:l3} compares the data and the
predictions for the ten centre-of-mass-averaged cross sections. 

The results of the study are expressed by inserting an additional
parameter, $C$, in the description of the
running of $\alpha$ of Equation~\ref{eq:alfa}:
\begin{equation}\label{eq:runC}
  \alpha(Q^{2}) = \frac{\alpha_0}{1-C\Delta\alpha(Q^{2})}.
\end{equation}
A fit to the 80 data points, which takes into account the dependence of the
measured integrated luminosity on $C$ shows that the value $C=0$,
corresponding to absence of running, is completely excluded. The data
are in excellent agreement with the running predicted in QED,
corresponding to $C=1$, and the fit yields:
\begin{equation}\label{eq:c}
C= 1.05 \pm 0.07 \pm 0.14,
\end{equation}
where the first uncertainty is statistical and the second
systematic. The systematic uncertainty is dominated by theoretical
uncertainties on the prediction of the Bhabha scattering differential
cross section both at large angles and in the luminosity monitor. Some
additional systematic contributions also come from the modelling of
the detector response.

\section{COMBINED RESULTS}

\begin{figure}
      \includegraphics[width=0.4\textwidth]{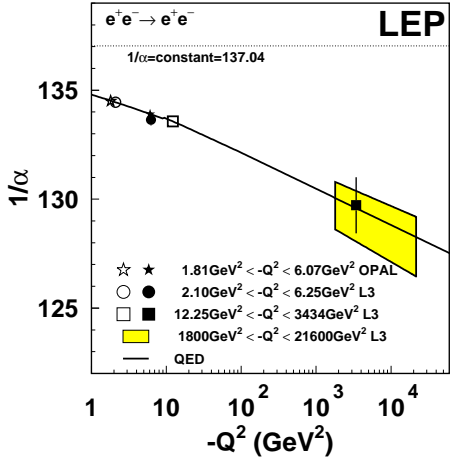}
    \caption{Summary of LEP results on the measurement of the
    running of the electromagnetic coupling. The  band
    represents the L3 measurement at high $Q^2$. The full symbols
    represent the OPAL and the L3 measurements at low and
    intermediate $Q^2$. The open symbols are the reference values to
    which the measurement are anchored, as discussed in the last
    section of the text.  The solid line shows the QED predictions of
      Reference~\cite{ref:burkhardt_new}.}
    \label{fig:7}
\end{figure}

Figure~\ref{fig:7}
summarises the LEP results on the
running of the electromagnetic coupling. The L3
measurement at $ 1800 \GeV^{2} < -Q^{2} < 21600
\GeV^{2}$~\cite{ref:l3} is represented as a  band, obtained by
inserting the measured value of $C$ from Equation~\ref{eq:c} into
Equation~\ref{eq:runC} and assuming the QED description of
$\Delta\alpha(Q^2)$ given in Reference~\cite{ref:burkhardt_new}.
The two low-$Q^2$ measurements by the OPAL and L3 collaborations at 
$1.8\GeV^2<-Q^2<6.1\GeV^2$ (Equation~\ref{eq:opal}) and
$2.1\GeV^2<-Q^2<6.2\GeV^2$ (Equation~\ref{eq:l3old1}), respectively, as well as
the intermediate-$Q^2$ measurement of L3 at
$12.25\GeV^2<-Q^2<3434\GeV^2$ (Equation~\ref{eq:l3old2}) are represented with two symbols each.
The empty symbols represent the values of
$\alpha^{-1}(Q^2)$ at the lower end of each $Q^2$ range. They are
fixed by using
Equation~\ref{eq:alfa} and the QED description of $\Delta\alpha(Q^2)$
of Reference~\cite{ref:burkhardt_new}. The full symbols represent the
values of $\alpha^{-1}(Q^2)$ at the higher end of each $Q^2$ range
extracted from the fixed values and from the measurements in
Equations~\ref{eq:opal},~\ref{eq:l3old1} and~\ref{eq:l3old2}. All
measurements are in excellent agreement with the QED predictions of
Reference~\cite{ref:burkhardt_new}, presented as a solid line.

\begin{figure}[h]
      \includegraphics[width=0.4\textwidth]{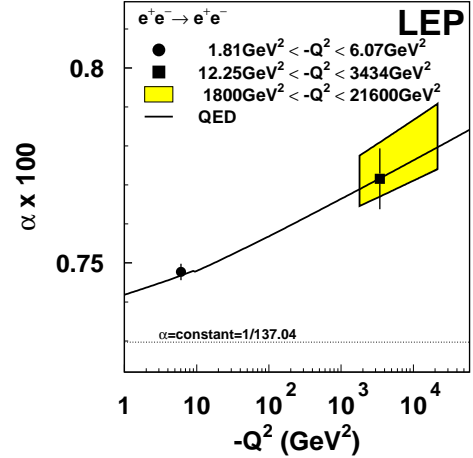}\\
    \caption{Combined LEP results on the measurement of the running of
      the electromagnetic coupling in three different $Q^2$ regimes,
      compared with the QED predictions of
      Reference~\cite{ref:burkhardt_new}. The treatment of data is
      discussed in the last section of the text.}
    \label{fig:8}
\end{figure}

Figure~\ref{fig:8} presents a combination of all LEP results. This
combination is obtained in
several steps. First, the L3 measurement
 at $2.1\GeV^2<-Q^2<6.2\GeV^2$ and the OPAL measurement at
 $1.8\GeV^2<-Q^2<6.1\GeV^2$, from Equations~\ref{eq:opal}
 and~\ref{eq:l3old1} respectively,
are  combined into a single measurement. In order to perform this
combination the L3 measurement is transported to 
the $Q^2$ range of the OPAL measurement. The combined result is:
\begin{equation}\label{eq:combi}
 \alpha(-6.1\GeV^2)-\alpha(-1.8\GeV^2)=(363\pm 52)\times 10^{-7},
\end{equation}
where the uncertainty combines statistical and systematic effects.
This combined result and the L3 measurement at intermediate $Q^2$ are
plotted in Figure~\ref{fig:8} as two single points at $Q^2=-6.1\GeV^2$
and  $Q^2=-3434 \GeV^{2}$, respectively. The values of $\alpha(Q^2)$ at these two
points are extracted by
 anchoring the value of $\alpha(Q^2)$ at the lower end of
 each $Q^2$ range by using the L3 measurement of $C$ at $ 1800
 \GeV^{2} < -Q^{2} < 21600 \GeV^{2}$ and assuming it also describes
 the running of $\alpha$ for lower values of $Q^2$:
the
 values of $\alpha(-1.8\GeV^2)$ and $\alpha(-12.25 \GeV^{2})$ are
fixed by using the measured
 value of $C$ from Equation~\ref{eq:c}, the evolution expected from
 Equation~\ref{eq:runC} and the QED description of $\Delta\alpha(Q^2)$
 of Reference~\cite{ref:burkhardt_new}. The value of
 $\alpha(-6.1\GeV^2)$ is finally extracted by using this fixed value of
 $\alpha(-1.8\GeV^2)$ and Equation~\ref{eq:combi}, with an additional uncertainty which follows
 from the 14\% uncertainty on $C$. A similar procedure is followed to
 extract the value of $\alpha(-3434 \GeV^{2})$ from
 Equation~\ref{eq:l3old2}.

In conclusion, the LEP experiments have established the 
the evolution of the electromagnetic coupling with the
squared four-momentum transfer in a new energy domain. These measurements, combined
in Figure~\ref{fig:8}, span three orders of magnitude in $Q^2$ and
confirm the QED predictions for the running of $\alpha$.

\end{document}